\documentclass[twocolumn,showpacs,amsmath,amssymb,pre]{revtex4}
\usepackage{graphicx}
\usepackage{soul}
\usepackage{color}
\usepackage{times}

\begin{document}
\title{
Emergence and control of heat current from strict zero thermal
bias}

\author{Jie Ren}
\email{renjie@nus.edu.sg}
\author{Baowen Li}
\email{phylibw@nus.edu.sg} 
\affiliation{NUS Graduate School for Integrative Sciences and Engineering, Singapore 117456, Republic of Singapore\\
         Department of Physics and Centre for Computational Science
         and Engineering, National University of Singapore, Singapore 117546, Republic of Singapore
         }

\begin{abstract}
It is an ever-growing challenge to develop nano-devices for
controlling energy transport. An open question is whether we can
create and control heat current at {\it strict zero thermal bias},
and if yes, how to do it. In this paper, we demonstrate that a
nonlinear asymmetric system, when pushed out of equilibrium, can
produce heat current in the absence of a thermal bias. The
emergence and control of heat current over a broad range of
parameters are studied. Our results reveal the following three
necessary conditions: non-equilibrium source, symmetry breaking,
and nonlinearity. We also demonstrate that when heat baths are
correlated, symmetry breaking is sufficient to generate heat
current.


\end{abstract}


\maketitle

Understanding heat transfer at the molecular level is of
fundamental and practical importance \cite{nano}. Recent years
have witnessed a fast development in the emerging field of
\emph{phononics} \cite{WL08}, wherein phonons, rather than an
annoyance, can be used to carry and process information. To
manipulate and control phonon transport (heat current) on the
molecular level, various
thermal devices \cite{WL08, memory} have been proposed. 
On the other hand, experimental works such as thermal rectifier
\cite{exp_diode} and nanotube phonon waveguide \cite{waveguide}
have been carried out. These theoretical and experimental works
render the heat current to be controlled as flexibly as electric
current in a foreseeable future.

Heat transfers spontaneously from a high temperature to a low one;
thus, the control of heat current has been so far based on the
control of the temperature gradient. However, a large temperature
gradient is essentially difficult to maintain over small distance
in practice, especially at nanoscale. Consequently, a natural
question is raised: can we create and control heat current in the
absence of (or against) thermal bias at nanoscale; if yes, then
how do we do that?

Inspired by ideas from Brownian motors \cite{brownianmotor},
originally devised for particle transport, a few studies have
revealed the possibility of pumping heat against thermal gradients
at nanoscale \cite{Broeck0608, Segal0608, Dhar07, Li0809}. A
molecular model with modulated energy levels has been found to
perform the heat pumping operation \cite{Segal0608}. While the
microscopic oscillator system, though built on the similar
principles, fails to perform the pumping \cite{Dhar07}. Thus, it
is still not clear what the requirements are for the system to
show such functional effect. In this paper, we attempt to answer
this novel and important question: how do we create and control
heat current at {\it strict zero thermal bias}?

It is noted that some interesting works reveal that nonzero heat
current survives when one bath temperature is driven but with
equal average (but different at any instant) to the other bath
temperature \cite{Li0809}. However, this reported behavior can be
understood through the Landauer formula for the heat current
\cite{Rego98}: $J=\int d\omega\omega
\mathcal{T}(\omega)[\eta(\omega, T_L)-\eta(\omega, T_R)]$, where
$\mathcal{T}(\omega)$ is the transmission coefficient and $\eta(
\omega, T_{L/R})$ is the Bose-Einstein distribution. Considering
the temperatures $T_{L/R}$ are driven around the same average
$T_0$, Taylor expansion gives: $\eta(\omega, T_L)-\eta(\omega,
T_R)\simeq\eta'(\omega,
T_0)[T_L(t)-T_R(t)]+\eta''[T_L(t)-T_R(t)]^2/2$. After the periodic
average, the first term vanishes while the second order survives
which produces the nonzero current.

Therefore, in a stark contract to the above proposals, we keep
{\it strict zero thermal bias at every instant} through our
studies. This seems to be a small step, but it is a revolutionary
one and has a completely different physics. It is under this
strict zero thermal bias that our results uncover these three
following conditions for the emergence of heat current at zero
thermal bias: non-equilibrium source, symmetry breaking, and
nonlinearity. Moreover, our simulation and analytic results reveal
a phenomenon that symmetry breaking is already sufficient, if the
two heat baths are correlated.

\begin{figure}
\scalebox{0.40}[0.40]{\includegraphics{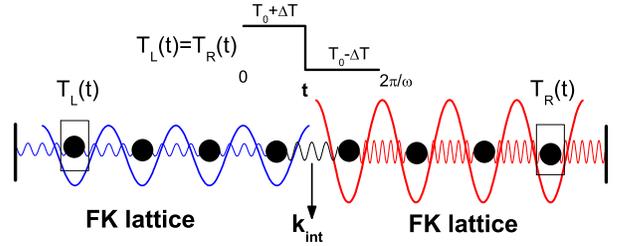}} \vspace{-.2cm}
\caption{(Color online). Schematic illustration of one-dimensional
two segment FK lattice, being coupled to two isothermal baths with
oscillating temperatures $T_L(t)=T_R(t)$.} \label{fig:1}
\end{figure}

Our system consists of two segment Frenkel-Kontorova (FK) chains
\cite{FK1, FK2} coupled together by a harmonic spring with
constant strength $k_{int}$ as depicted in Fig. \ref{fig:1}. The
Hamiltonian can be written as:
\begin{eqnarray}
H=H_L+\frac{k_{int}}{2}(q_{N_L,L}-q_{1,R})^2+H_R,
\end{eqnarray}
 where the Hamiltonian of each FK segment reads:
\begin{eqnarray}
H_S=\sum_{i=1}^{N_S}
\frac{p^2_{i,S}}{2m}+\frac{k_S}{2}(q_{i,S}-q_{i+1,S})^2
-\frac{V_S}{(2\pi)^2}\cos\frac{2\pi q_{i,S}}{a}.
\end{eqnarray}
 $S$ stands for $L$ or $R$, which represents the left or right
segment with the same length. $q_{i,S}$ denotes the displacement
from the equilibrium position for the $i$th atom in segment $S$
and $p_{i,S}$ the corresponding momentum. $a$ is the lattice
constant. $k_S$ and $V_S$ are the spring constant and the strength
of the on-site potential of segment $S$. Two isothermal baths
contacted with two ends are simulated by Langevin reservoirs with
zero mean and variance $\langle \xi_{1/N}(t) \xi_{1/N}(t') \rangle
= 2\gamma k_BT_{L/R}\delta(t-t')$, where $\gamma$ is the
system-bath
coupling strength. 

\begin{figure}
\scalebox{0.64}[0.64]{\includegraphics{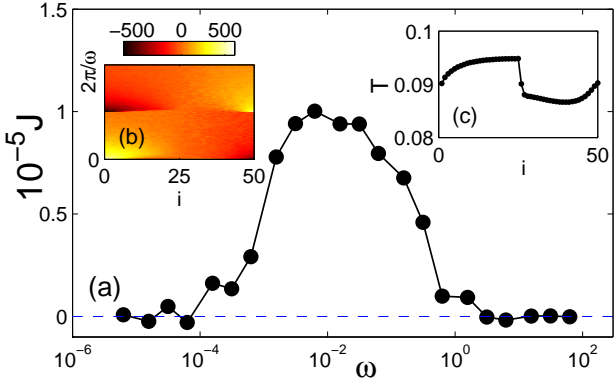}} \vspace{-.5cm}
\caption{(Color online). {\bf Frequency resonance effect}. (a) $J$
versus $\omega$.  (b) The cyclic local heat current
$\overline{J}_i(t)$ (see definition in context) is illustrated at
the optimal driving frequency $\omega=2\pi\times10^{-3}$. (c)
$T_i$ profile is shown for the same $\omega$. $N=50, T_0=0.09,
\Delta T=0.045, 20k_{int}=5k_L=k_R=1, 5V_L=V_R=5, \gamma=0.5$,
throughout the paper, except specified.} \label{fig:2}
\end{figure}

One question of interest is whether, due to the spatial asymmetry
of the system, the thermal fluctuation in heat baths can induce a
net heat current in a given direction. We argue that it is
impossible. The situation would be a perpetual machine of the
second kind extracting useful work out of ambient thermal
reservoirs of vast energy surrounding us. Unfortunately, the
second law of thermodynamics rules out the hiding place of the
Maxwell demon, no matter how smart you design the system. It seems
that any design to generate a heat current without thermal bias is
foolish and even a quackery in the face of the second law.
However, the second law works at thermal equilibrium only.

In this paper, we drive the system out of equilibrium by
periodically oscillating two isothermal baths simultaneously, as
$T_L(t)=T_R(t)=T_0+\Delta T {\rm sgn}(\sin\omega t)$, where $T_0$
is the reference temperature. Under the time-varying heat baths,
in the long-time limit, the local temperature of site $i$ is time
periodic, namely, $T_i(t)= m
\dot{q}^2_i(t)/k_B=T_i(t+2\pi/\omega)$. Similarly, the
time-dependent local heat current has the same periodicity:
$J_i(t)=k\dot q_i(t)(q_i(t)-q_{i+1}(t))=J_i(t+2\pi/\omega)$.
Therefore, within one period $t\in(0, 2\pi/\omega)$, we can define
the cyclic local heat current averaged over the ensemble of
periods after the transient time:
$\overline{J}_i(t)=\frac{1}{n}\sum^n_{k=1} J_i(t+2k\pi/\omega)$
($n$ is the number of periods) as well as the cyclic local
temperature $\overline{T}_i(t)= \frac{1}{n}\sum^n_{k=1}
T_i(t+2k\pi/\omega)$. Thus, the net heat current and the effective
local temperature read
$J=\frac{\omega}{2\pi}\int^{\frac{2\pi}{\omega}}_{0}\overline{J}_i(t)
dt$ and $T_i=\frac{\omega}{2\pi}\int^{\frac{2\pi}{\omega}}_{0}
\overline{T}_i(t) dt,$ which are the same as the long-time
average.

For convenience of numerical calculations, we use dimensionless
parameters by measuring positions in units of $[a]$, momenta in
units of $[ a(mk_R)^{1/2}]$, spring constants in units of $[k_R]$,
frequencies in units of $[(k_R/m)^{1/2}]$, and temperatures in
units of $[a^2k_R/k_B]$. A fixed boundary condition is applied and
the equation of motion is integrated by the symplectic velocity
Verlet algorithm with a time step of $0.005$ for a sufficiently
long time to guarantee the nonequilibrium stably periodic states.

\begin{figure}
\scalebox{0.34}[0.36]{\includegraphics{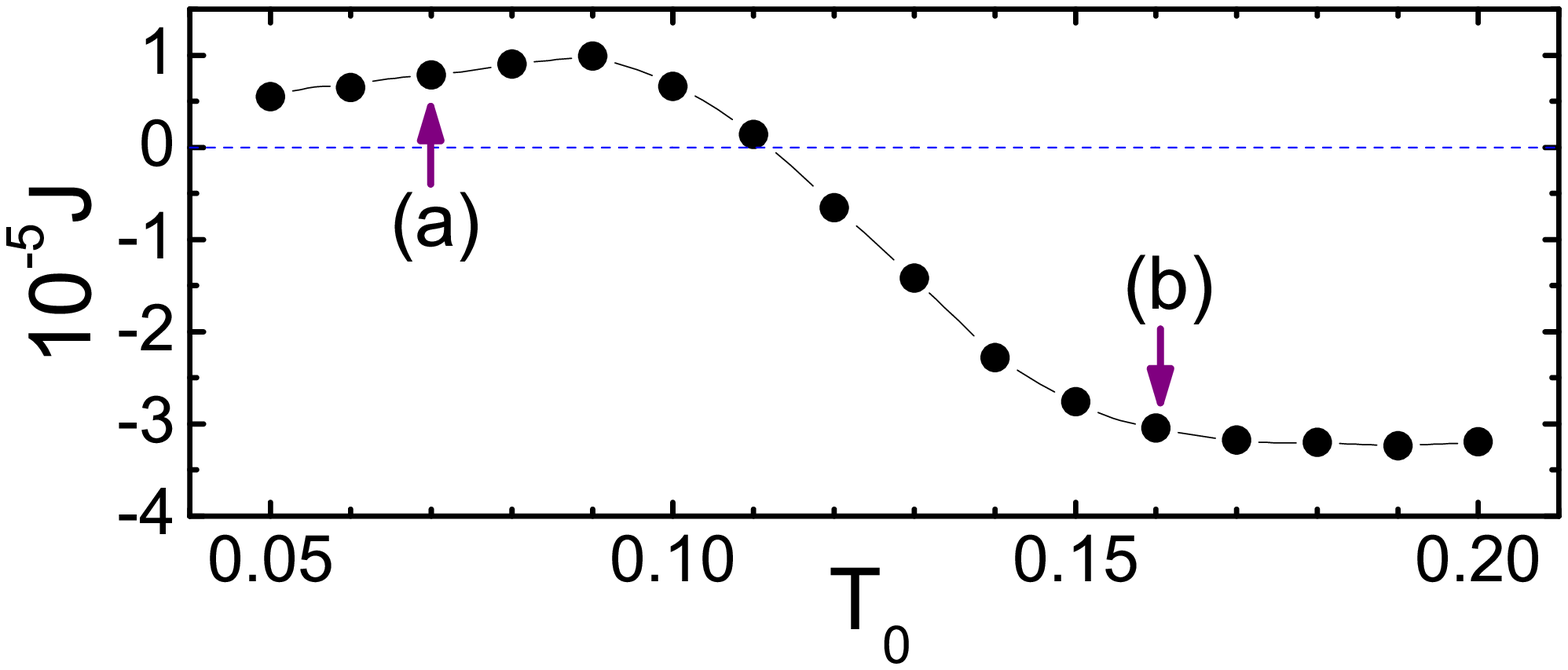}}\vspace{-.1cm}
\scalebox{0.6}[0.63]{\includegraphics{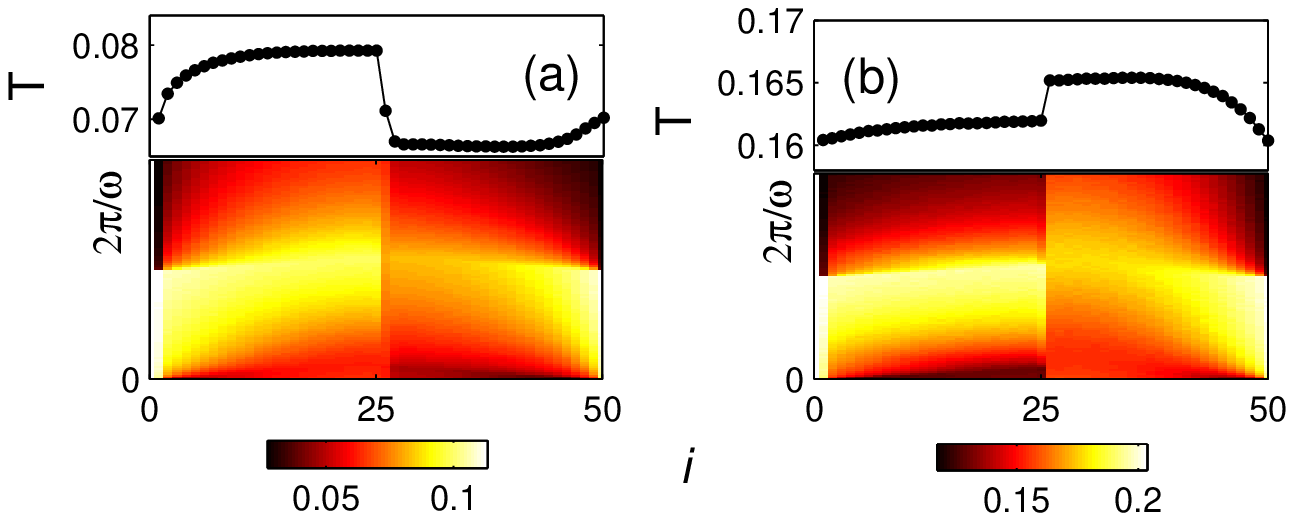}} \vspace{-.3cm}
\caption{ (color online). {\bf Temperature tuning effect} (upper
panel). $\omega=2\pi\times10^{-3}$. The middle and bottom panels
show $T_i$ profiles and $\overline{T}_i(t)$ patterns with $T_0=$
$0.07$ (a) and $0.16$ (b). At low temperature case (a), the left
segment has lower conductivity which results in slower heat
dissipating. Thus, more energies cumulate at the left part, making
$J$ from left to right. While at high temperature case (b), the
right part has lower conductivity which induces the reversal of
$J$.} \label{fig:3}
\end{figure}

By adjusting the driving frequency $\omega$ of the two isothermal
baths simultaneously, we obtain a \emph{nonzero} heat current,
which can be maximized at a moderate $\omega$, as shown in Fig.
\ref{fig:2}a. It is intuitive that the heat current vanishes at
both high and low frequency regimes. In the fast-oscillating limit
$\omega\rightarrow\infty$, thermal baths are driven so fast that
two ends of the lattice cannot respond accordingly. The system
only feels the same time-averaged temperature $T_0$ at the ends,
which yields $J\rightarrow0$. In the adiabatic limit
$\omega\rightarrow0$, the system reduces to its equilibrium
counterpart without thermal gradients so that $J\rightarrow0$. The
emergence of heat current happens only when two segments of the
system respond differently.

The $\overline{J}_i(t)$ pattern at the optimized frequency is
shown in Fig. \ref{fig:2}b. At the first half driving period with
positive temperature variation, the heat transports from the two
ends to the central part. While at the second half driving period
with negative temperature variation, the direction of heat current
reverses from the central part to the two ends. Eventually, the
net heat current emerges due to the asymmetry of heat conduction
resulting from the asymmetrical segment structure. Moreover, we
illustrate the $T_i$ profile in Fig. \ref{fig:2}c. It indicates
that more energies are cumulated at the left segment part which in
turn induces the net heat current from left to right although two
heat baths hold the same temperature. As we mentioned, the
emergence of heat current here does not break down the second law
of thermodynamics since the system is pushed out of equilibrium.
The only non-equilibrium source driving the heat current is the
oscillating temperature $T_L(t)$ and $T_R(t)$ other than the
thermal gradients. This phenomenon is somehow similar to that
resonance in the particle current observed in the
temperature ratchet \cite{Peter96}.

The reference temperature $T_0$ is a very important parameter
since thermal conductivity is generally temperature dependent in
many nonlinear lattices \cite{Li07}. In the upper panel of Fig.
\ref{fig:3}, we show that the direction of the net heat current
reverses as the reference temperature $T_0$ increases and then
saturates at high temperatures. To gain more insights into this
reversal phenomenon, we depict the effective local temperature
profile $T_i$ and the cyclic local temperature pattern
$\overline{T}_i(t)$ with two typical values of $T_0$ in the middle
and lower panels of Fig. \ref{fig:3}. It shows clearly that at low
temperatures, energy dissipates faster at the right segment with
high thermal conductivity which makes heat cumulated at the left
part. Thus, there is a net heat current flowing from left to
right. While at high temperatures, the scenario is reversed. In
other words, the heat current flows from the segment with lower
conductivity to the higher one and the reversal of heat currents
results from the order reversal of thermal conductivities of two
FK segments as temperature increases. Further, we check the
amplitude tuning effect by varying $\Delta T$. As expected, it is
found that the larger the $\Delta T$, the larger the magnitude of
$J$.

\begin{figure}
\scalebox{0.42}[0.42]{\includegraphics{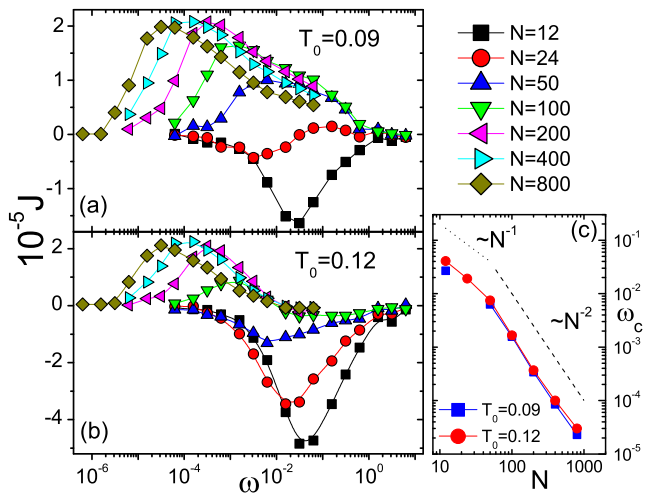}} \vspace{-.2cm}
\caption{(Color online). {\bf System size effect.} $J$ versus
$\omega$ with various sizes $N$ for two different reference
temperatures $T_0$: (a) 0.09; (b) 0.12, with $\Delta T=T_0/2$. (c)
$\omega_c$ versus $N$ indicates the ballistic transport and the
normal diffusion.} \label{fig:4}
\end{figure}

By varying $N$, we find that the maximum value of $J$ increases
with system size and then saturates as shown in Fig. \ref{fig:4}.
Moreover, the optimum frequency $\omega_c$ decreases as $N$
increases. This ``redshift" can be understood from thermal
response time \cite{Li0809}. The heat conduction in the FK lattice
follows Fourier's law when the system size is larger enough
\cite{FK2}, thus, the response time, $\tau\sim N^2$,
characterizing the time scale for the energy to diffuse across the
system. Therefore, the characteristic frequency scales as
$\omega_c\propto N^{-2}$, which explains the redshift of the
optimum frequency $\omega_c$ when $N$ increases. More
interestingly, we find that when $N$ is decreased, the direction
of $J$ can be reversed and then the magnitude of $J$ increases
although negative. The optimum frequency $\omega_c$ in this small
size regime, scales as $N^{-1}$ instead of the scaling $ N^{-2}$
of the normal diffusion as shown in Fig. \ref{fig:4}c. This is
because at small $N$ regime, the phonon transports ballistically.
Since the right segment is more rigid than the left one
($k_R=5k_L$), the energy transports faster in the right part which
induces $J$ from right to left so as to reverse the direction. In
other words, the crossover from normal diffusion to ballistic
transport owing to the size reducing induces the reversal of the
net heat current .

\begin{figure}
\scalebox{0.42}[0.43]{\includegraphics{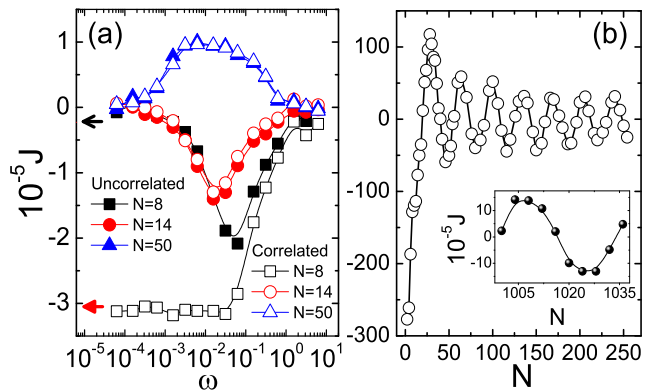}} \vspace{-.4cm}
\caption{(Color online). (a) {\bf Thermal bath correlation
effect}. $c=2\gamma k_BT_{L/R}(t)$. The upper black arrow marks
$J$, without driving, under correlated baths with $N=8$; while the
bottom red arrow marks $J$ in the adiabatic limit, with the same
condition. Their discrepancy indicates the geometric phase induced
heat current. (b) Correlation-induced $J$ versus $N$. Results are
calculated from the analytic Eq. (\ref{RLL}) and are checked that
they are the same as the simulation results. $\Delta T=V_L=V_R=0,
c=2\gamma k_BT_0$.} \label{fig:5}
\end{figure}

In all simulations so far, two isothermal baths are independent,
since $\langle \xi_1(t) \xi_N(t') \rangle =0$. Now we introduce
the nonzero variance $\langle \xi_1(t) \xi_N(t')
\rangle=c\delta(t-t')$ to study the thermal bath correlation
effect. We find that the correlation effect is significant only at
small sizes (in which the energy transports ballistically) while
fading away at large sizes in which energy transports diffusively,
as shown in Fig. \ref{fig:5}a. When $N=8$, we even obtain a
nonzero heat current in the adiabatic limit $\omega\rightarrow0$.
It implies that there may exist a``geometric phase" in the
dissipative and stochastic system \cite{phase} we used here. The
geometric phase results from the nonzero area enclosed by periodic
variation of parameters in parameter space. Even when the driving
is extremely slow, approximating to equilibrium state at every
instant, this geometric phase-induced heat current still survives.
This is an interesting topic deserving further investigation.

Moreover, in contrast to the nonlinear FK lattice, we find
surprisingly that, for a pure harmonic system, even in the absence
of external driving, and \emph{without} thermal bias, the heat
current emerges when the two thermal baths are correlated. To
understand this correlation-induced heat current, we formalize the
heat conduction of harmonic systems by the Rieder-Lebowitz-Lieb
method \cite{RLL} and have the steady-state equation:
\begin{equation}\label{RLL}
\hat{A}\cdot\hat{B}+\hat{B}\cdot\hat{A}^T=\hat{D},
\end{equation}
where
$\hat{A}=\left(%
\begin{array}{cc}
  0 & -\hat{I} \\
  \hat{L} & \hat{\Gamma} \\
\end{array}%
\right),
 \hat{D}=\left(%
\begin{array}{cc}
  0 & 0 \\
  0 & \hat{C} \\
\end{array}%
\right). $ $L_{ij}=\delta_{ij}\sum_{m}k_{im}-k_{ij}$ is the
Laplacian matrix where $k_{ij}$ denotes the spring constant
between adjacent oscillator $i$ and $j$.
$\Gamma_{in}=\gamma_i\delta_{in}, (n=1,N)$ is the dissipation
matrix.
$C_{ij}=2\gamma_ik_BT_i(\delta_{i1}\delta_{j1}+\delta_{iN}\delta_{jN})
+c(\delta_{i1}\delta_{jN}+\delta_{iN}\delta_{j1})$ and the second
term depicts the correlation effect. The solution $\hat{B}$ of Eq.
(\ref{RLL})
 has four blocks:
$\hat{B}=\left(
\begin{array}{cc}
  \hat{B}_{qq} & \hat{B}_{qp} \\
  \hat{B}^T_{qp} & \hat{B}_{pp} \\
\end{array}
\right), $ where $(\hat{B}_{qp})_{ij}=\langle q_ip_j\rangle$
denotes the position-velocity correlation relating to $J$. For two
oscillator system, it is easy to obtain the explicit solution of
heat currents:
\begin{equation}
J=\frac{k_{int}[(k_L-k_R)c+2\gamma(T_L-T_R)k_{int}]}{(k_L-k_R)^2+2\gamma^2(k_L+k_R)+4\gamma^2k_{int}+4k_{int}^2}.
\end{equation}
It shows clearly that even \emph{without} thermal bias
($T_L=T_R$), the correlation $c$ still can induce nonzero $J$ in
the asymmetry harmonic chain ($k_L\neq k_R$). This nontrivial
correlation effect, represented by the nonzero off-diagonal of
$C_{ij}$, is reminiscent of the quantum coherence \cite{Single00,
Scully03}, which provides the off-equilibrium source to create
heat currents. More interestingly, we find that when $N$ increases
the correlation-induced heat current oscillates and reverses its
direction periodically, as shown in Fig. \ref{fig:5}b.

In summary, we have demonstrated the emergence of heat currents
and their direction control in two segments FK lattice
\emph{without} thermal bias. By periodically oscillating two
isothermal baths simultaneously, we can obtain a maximum heat
current at an optimum driving frequency. The resonance effect with
respect to other parameters has been studied systematically as
well. We have found that the direction of the emerging heat
current can be reversed by either tuning the reference temperature
or tuning the system size, and the optimum frequency can be
decreased by increasing the system size. Our results reveal the
following three necessary conditions for the emergence of heat
current \emph{without} thermal bias: (1) nonequilibrium source,
which is induced by the periodically oscillating baths although
isothermal, and in turn breaks the underlying detailed balance.
Also, we can periodically drive other parameters such as $k_{int}$
to generate the nonequilibrium source so as to create heat current
(not shown); (2) symmetry breaking, which results from the two
asymmetric segments construction defining a preferential
directionality; (3) nonlinearity, which comes from the on-site
sinusoidal potential in the FK model. In fact, when $V_{L/R}=0$,
the system reduces to an asymmetric harmonic chain and we find
that the heat current vanishes (not shown). However, if the
correlation between two baths is introduced, symmetry breaking is
sufficient to create heat current.

We should point out that the model proposed here might be realized
experimentally. For a typical atom, $a\sim1$ \AA,
$m\sim10^{-26}-10^{-27}$ kg, which yields the frequency unit
$[(k_R/m)^{1/2}]\sim10^{13}$ s$^{-1}$ and the temperature unit
$[a^2k_R/k_B]\sim10^3-10^4$ K. The typical value of $T_0=0.1$ and
$\omega\sim10^{-4}-10^{-5}$ corresponds to the physical
temperature $T_r\sim10^2-10^3$ K and the physical frequency
$\omega_r\sim10^2-10^3$ MHz, which is in the ultrasonic and
microwave regimes. Thanks to the redshift effect, we are able to
obtain lower optimum driving frequencies in practice by enlarging
the system size. The thermal bath correlation might be implemented
by imposing some common external thermal noise, or introducing
entanglements between two heat baths \cite{Scully03}. We hope the
present study will stimulate experimentalists to search possible
realizations and technological utilizations.

We thank Nianbei Li, Lifa Zhang and Peter~H\"anggi for fruitful
discussions. This work was supported in part by ARF Grant No.
R-144-000-203-112 from the Ministry of Education of the Republic
of Singapore and Grant No. R-144-000-222-646 from the National
University of Singapore.

\end{document}